\documentclass[RNAAS]{aastex631}

\usepackage{mathtools}

\shorttitle{ASASSN-21co}
\shortauthors{Rowan et al.}

\begin{document}

\title{ASASSN-21co: A detached eclipsing binary with an 11.9 year period}

\author[0000-0003-2431-981X]{D. M. Rowan}
\affiliation{Department of Astronomy, The Ohio State University, 140 West 18th Avenue, Columbus, OH, 43210, USA}
\affiliation{Center for Cosmology and Astroparticle Physics, The Ohio State University, 191 W. Woodruff Avenue, Columbus, OH, 43210, USA}

\author{K. Z. Stanek}
\affiliation{Department of Astronomy, The Ohio State University, 140 West 18th Avenue, Columbus, OH, 43210, USA}
\affiliation{Center for Cosmology and Astroparticle Physics, The Ohio State University, 191 W. Woodruff Avenue, Columbus, OH, 43210, USA}

\author{Z. Way}
\affiliation{Department of Astronomy, The Ohio State University, 140 West 18th Avenue, Columbus, OH, 43210, USA}
\affiliation{Center for Cosmology and Astroparticle Physics, The Ohio State University, 191 W. Woodruff Avenue, Columbus, OH, 43210, USA}

\author{C. S. Kochanek}
\affiliation{Department of Astronomy, The Ohio State University, 140 West 18th Avenue, Columbus, OH, 43210, USA}
\affiliation{Center for Cosmology and Astroparticle Physics, The Ohio State University, 191 W. Woodruff Avenue, Columbus, OH, 43210, USA}

\author{T. Jayasinghe}
\affiliation{Department of Astronomy, The Ohio State University, 140 West 18th Avenue, Columbus, OH, 43210, USA}
\affiliation{Center for Cosmology and Astroparticle Physics, The Ohio State University, 191 W. Woodruff Avenue, Columbus, OH, 43210, USA}

\author{Todd A. Thompson}
\affiliation{Department of Astronomy, The Ohio State University, 140 West 18th Avenue, Columbus, OH, 43210, USA}
\affiliation{Center for Cosmology and Astroparticle Physics, The Ohio State University, 191 W. Woodruff Avenue, Columbus, OH, 43210, USA}

\author{H. Barker}
\affiliation{Rutherford Street Observatory, Nelson, New Zealand}

\author{F.-J. Hambsch}
\affiliation{Vereniging Voor Sterrenkunde (VVS), Oostmeers 122 C, 8000 Brugge, Belgium}
\affiliation{Bundesdeutsche Arbeitsgemeinschaft für Veränderliche Sterne e.V. (BAV), Munsterdamm 90, D-12169 Berlin, Germany}
\affiliation{American Association of Variable Star Observers (AAVSO), 49 Bay State Road, Cambridge, MA 02138, USA}

\author{T. Bohlsen}
\affiliation{Mirranook Observatory, Armidale NSW Aust}

\author{Stella Kafka}
\affiliation{American Association of Variable Star Observers (AAVSO), 49 Bay State Road, Cambridge, MA 02138, USA}

\author{B. J. Shappee}
\affiliation{Institute for Astronomy, University of Hawaii, 2680 Woodlawn Drive, Honolulu, HI 96822, USA}

\author{T. W. -S. Holoien}
\affiliation{Carnegie Observatories, 813 Santa Barbara Street, Pasadena, CA 91101, USA}

\author{J. L. Prieto}
\affiliation{N\'ucleo de Astronom\'ia de la Facultad de Ingenier\'ia y Ciencias, Universidad Diego Portales, Av. Ej\'ercito 441, Santiago, Chile}
\affiliation{Millennium Institute of Astrophysics, Santiago, Chile}

\correspondingauthor{D. M. Rowan}
\email{rowan.90@osu.edu}
\begin{abstract}
    We use ASAS $V$-band and ASAS-SN $g$-band observations to model the long-period detached eclipsing binary ASASSN-21co. ASAS observations show an eclipse of depth $V\sim0.6$ mag in April of 2009. ASAS-SN $g$-band observations from March of 2021 show an eclipse of similar duration and depth, suggesting an orbital period of 11.9 years. We combine the $g$-band observations with additional BVRI photometry taken during the eclipse to model the eclipse using PHOEBE. We find that the system is best described by two M giants with a ratio of secondary radius to primary radius of $\sim0.61$. Optical spectra taken during the eclipse are consistent with at least one component of the binary being an M giant, and we find no temporal changes in the spectral features. The eclipse itself is asymmetric, showing an increase in brightness near mid-eclipse, likely due to rotational variability that is too low amplitude to be observed out-of-eclipse.
\end{abstract}


\section{}

Detached eclipsing binaries present a unique opportunity to measure accurate masses and radii of stars that are effectively in isolation \citep[e.g., ][]{Torres10}. The All-Sky Automated Survey for SuperNovae \citep[ASAS-SN, ][]{Shappee14, Kochanek17} observes the entire sky in optical light in the $V$-band (between 2012 and mid-2018) and $g$-band (since 2018). Although primarily designed for transient phenomena, the $V$-band observations have been used to classify $\sim 426,000$ variables, including $\sim150,000$ eclipsing binaries \citep{Jayasinghe19II, Jayasinghe21IX}. The ongoing ASAS-SN Citizen Science campaign \citep{Christy21} has also already identified a host of new variable stars in the $g$-band data. 

\begin{figure}
    \centering
    \includegraphics[width=0.9\linewidth]{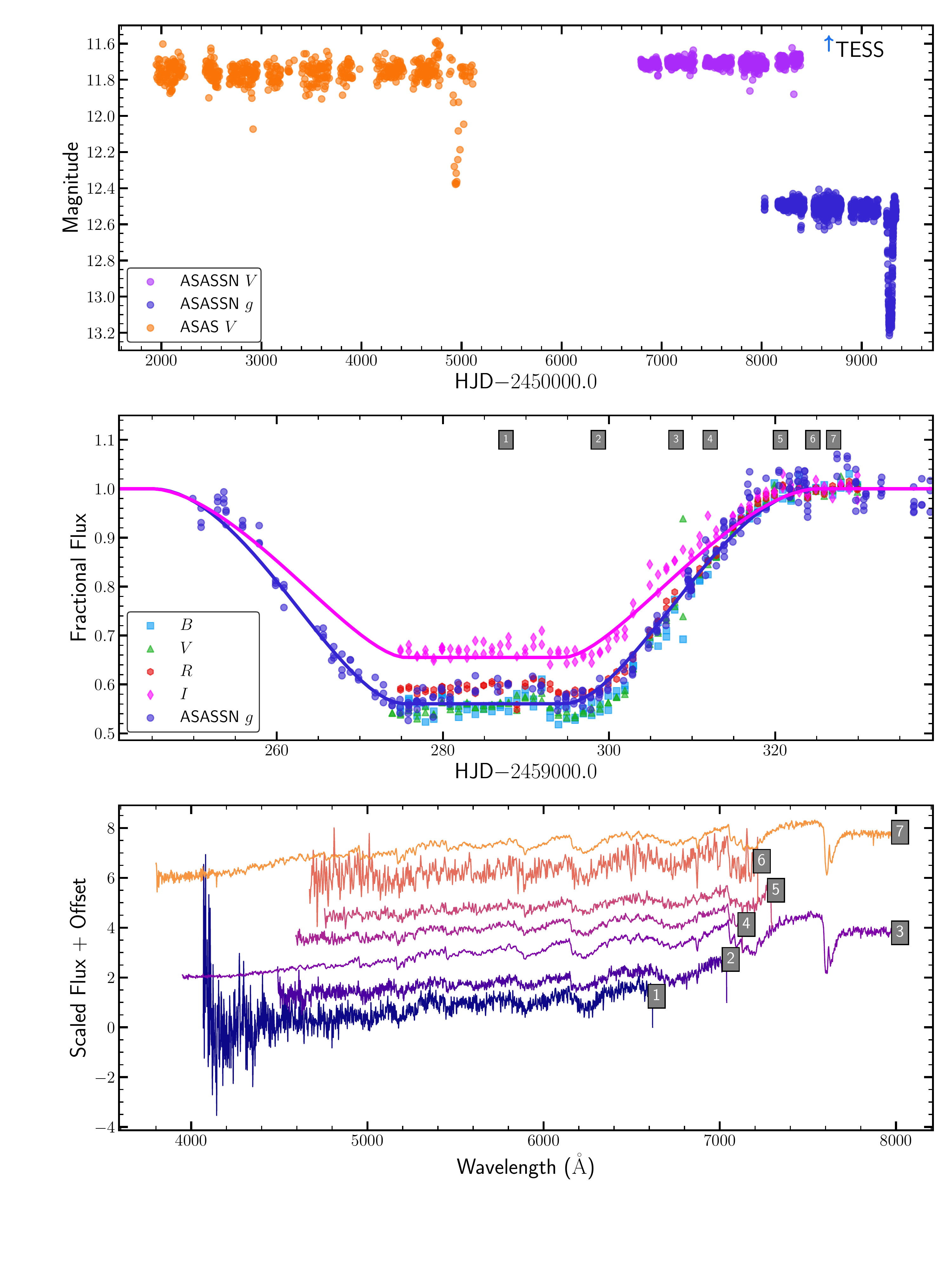}
    \caption{Top: ASAS and ASAS-SN light curves of 21co. The time of the TESS sector 13 is labeled by the blue arrow. We see a single eclipse in the ASAS $V$-band data and in the ASAS-SN $g$-band observations $\sim 11.9$ years later. Middle: the ASAS-SN $g$-band eclipse with additional BVRI measurements The {\tt PHOEBE} fits to the $g$- and $I$-band data are shown as solid lines. Bottom: scaled and shifted spectra of ASASSN-21co ordered in time from bottom to top. The orbital phase of each spectra is labeled in the middle panel with the corresponding number.}
    \label{fig:lc_spectra}
\end{figure}

Here we characterize the long-period detached eclipsing binary ASASSN-21co (RA=18:17:51.52, DEC$=-$~58:07:49.31). ASASSN-21co was first identified by \citet{Way21} as a long-period eclipsing binary with an eclipse depth of $g\sim0.6$~mag. $V$-band observations from the All-Sky Automated Survey \citep[ASAS, ][]{Pojmanski97} also show a similar eclipse around UT 2009-04-23. We use the {\tt astrobase} \citep{Bhatti20} implementation of the box least squares algorithm to derive an orbital period of $P=4344$~days (11.9 years). The absolute \textit{Gaia} magnitude \citep{GaiaCollaboration18}, corrected for extinction, is $M_G=-1.76$, with a \textit{Gaia} color of $G_{\rm{BP}}-G_{\rm{RP}}=2.23$~mag \citep{Anders19}, suggesting the star is a luminous red giant. The WISE and 2MASS colors, $W_1-W_2=-0.196$~mag and $J-K_s=1.093$~mag \citep{Curti12}, suggest that the photometric primary is an M giant \citep{Li16}. This is consistent with the spectra shown in Figure \ref{fig:lc_spectra}. 

Given the orbital period, the binary separation is $a\simeq5.2 ({M_T}/{M_\odot})^{1/3}\ \rm{AU}$, where $M_T$ is the total mass of the binary. The eclipse total duration is $t_T \sim66$~days and the time of total eclipse is $t_F \sim23$~days. If we assume that the impact parameter is zero, the radii of the primary $R_p$ and secondary $R_s$ are

\begin{equation} \label{eqn:radii}
    R_p = \frac{\pi (t_T+t_F)}{2P}a = 36.1\ R_\odot \left(\frac{M_T}{M_\odot}\right)^{1/3}\ \text{and}\ R_s = \frac{\pi (t_T - t_F)}{2P}a = 17.4\ R_\odot \left(\frac{M_T}{M_\odot}\right)^{1/3},
\end{equation}

\noindent The \citet{Anders19} {\tt StarHorse} catalog of \textit{Gaia} DR2 sources gives $M_*=1.06\ M_\odot$. Using \texttt{Isoclassify} \citep{Huber17,Berger20}, we combine the {\tt StarHorse} stellar parameters with MIST tables of bolometric corrections \citep{Dotter16} to find $R_p=73.26\ R_\odot$. Given this radius, Equation \ref{eqn:radii} implies a large total mass $M_T=8.4\ M_{\odot}$, suggesting that the {\tt StarHorse} radius is overestimated. 

Figure \ref{fig:lc_spectra} shows the ASAS and ASAS-SN $V$- and $g$-band observations. BVRI photometry taken during the 2021 eclipse shows that the eclipse is chromatic. Whereas the $g$-band magnitude falls by $\sim 0.6$, the $I$ magnitude decreases by only $\sim0.4$. The similarities between the asymmetric eclipse profile and disk eclipsing systems like $\epsilon$ Aurigae \citep{Kloppenborg10}, OGLE-LMC-ECL-11893 \citep{Dong14}, and ASASSN-V J192543.72$+$402619.0 \citep{Jayasinghe18detached} could suggest a dusty disk eclipse. This scenario would also produce a shallower eclipse in redder bands. In the case of a disk eclipse, the lack of secondary eclipse would be expected.

We use {\tt PHOEBE} \citep{Prsa16} to model the $g$- and $I$-band light curves of the most recent eclipse. We find that the eclipse depth and duration is consistent with two M giants with a ratio of radii $R_s/R_p=0.61$, close to our rough estimate of $0.48$ from the eclipse timing, and effective temperatures $T_{\rm{eff},s}/T_{\rm{eff},p}=1.09$. Since the redder bands have shallower eclipse depths, the observed eclipse occurs when the larger, colder primary is behind the smaller, hotter secondary. The eclipses observed in the ASAS and ASAS-SN data have the same duration and depth, while the eclipse structure implies that the secondary eclipse should have roughly half the depth of the primary eclipse, which rules out a period of $2P$. Shorter periods ($P/2$, $P/4$, etc.) have contradictions in their phased light curves because of missing eclipses. Finally, we note that the lack of a secondary eclipse near phase $0.5$ implies that the orbit must be elliptical. 

The eclipse profile is asymmetric, with larger variability seen in the ASAS-SN $g$-band data than the BVRI photometry. This could be due to low amplitude rotational variability on the cooler star. Using {\tt Period04} \citep{Lenz04}, we find low amplitude variability in the ASAS-SN $g$-band data at a period of 9.7~days. We find the same periodicity in the TESS sector 13 light curve from the Quick-Look-Pipeline \citep[QLP, ][]{Huang20a, Huang20b} in addition to larger amplitude periodicity at longer periods $\sim 20$~days. Since only one sector of TESS data is available, we cannot accurately constrain the period of this long-term variability. The longer period, higher amplitude variability is consistent with what is expected from overtone pulsators with higher modes \citep{Tabur10, Ferreira15}, while the shorter period may be either overtone pulsations or solar-type oscillations \citep{Mosser13}. 

Figure \ref{fig:lc_spectra} shows optical spectra taken from mid-eclipse to the completion of egress. The spectrum is consistent with that of an M giant and we see no temporal changes in the spectral features. The lack of spectral evolution during the eclipse is similar to the $\epsilon$ Aurigae system, whose spectrum does not vary with orbital phase despite the light curve suggesting a total eclipse \citep{Huang65}. The lack of spectral evolution does not rule out the two-giant scenario, as the effective temperature ratio found by the {\tt PHOEBE} model fits would also result in little spectral evolution during eclipse.

\clearpage

\bibliography{rowanbib}{}
\bibliographystyle{aasjournal}

\end{document}